\documentclass[10pt]{iopart}
\usepackage[UKenglish]{babel}
\usepackage{iopams}
\usepackage{setstack}
\usepackage{cite}
\usepackage{prettyref}
  \newcommand{\pr}[1]{\prettyref{#1}}
  \newrefformat{eqn}{(\ref{#1})}
  \newrefformat{sec}{section~\ref{#1}}
  \newrefformat{subsec}{section~\ref{#1}}
  \newrefformat{subsubsec}{section~\ref{#1}}
  \newrefformat{fig}{figure~\ref{#1}}
\usepackage{url}
\usepackage{amsfonts}
\usepackage{xspace}
\usepackage{dsfont}
\usepackage[dvips]{graphicx}
\usepackage{subfigure}
\eqnobysec
  \newcommand{\be}{\begin{equation}}
  \newcommand{\ee}{\end{equation}}
  \newcommand{\bea}{\begin{eqnarray}}
  \newcommand{\eea}{\end{eqnarray}}
\newcommand{\landau}{\mathcal{O}}
\newcommand{\p}{\partial}

\newcommand{\mc}[1]{\mathcal{#1}}
  \newcommand{\dint}[2]{\int \frac {d^{#2} #1}{ (2 \pi)^{#2} } \,}
\newcommand{\lb}{\overline{L}} 

\newcommand{\zb}{\zeta}
\newcommand{\ec}{\mathcal{E}_{C}}
\newcommand{\esc}{\mathcal{E}_{\rm sc}}
\begin{document}
\title{Casimir Energy of a {BEC}:
\\
From Moderate Interactions to the Ideal Gas}
\author{J Schiefele and  C Henkel}
\address{Institut f\"ur Physik und Astronomie, 
	Universit\"at Potsdam,
	Karl-Liebknecht-Str.~{24/25},
	{14\,476}~Potsdam,
	Germany}
\eads{\mailto{juergen.schiefele@physik.uni-potsdam.de}}
\begin{abstract}
Considering the Casimir effect due to phononic excitations of a weakly interacting dilute {BEC}, 
we derive a re-normalized expression for the zero temperature Casimir energy
$\ec$ of a {BEC}
confined to a parallel plate geometry with periodic boundary
conditions.  Our expression is formally equivalent to the free energy of
a bosonic field at finite temperature, with a nontrivial density of 
modes that we compute analytically.
As a function of the interaction strength, $\ec$ smoothly describes the
transition from the weakly interacting Bogoliubov regime to the
non-interacting ideal {BEC}.  For the weakly interacting case, $\ec$
reduces to leading order to the Casimir energy due to zero-point
fluctuations of massless phonon modes.  In the limit of an ideal
Bose gas, our result correctly describes the Casimir energy going to
zero.
\end{abstract}
%\submitto{\JPA}
%\maketitle
%
%
%%%%%%%%%%%%%%%%%%%%%%%%%%%%%%%%%%%%%%%%%%%%%%%%%%%%%%%%%%%%%%%%%%%%%%%%%%%%
% MAIN TEXT
%%%%%%%%%%%%%%%%%%%%%%%%%%%%%%%%%%%%%%%%%%%%%%%%%%%%%%%%%%%%%%%%%%%%%%%%%%%%
%
%
%%%%%%%%%%%%%%%%%%%%%%%%%%%%%%%%%%%%%%%%%%%%%%%%%%%%%%%%%%%%%%%%%%%%%%%%%%%%%%
\section{Introduction}
\label{sec:intro }
%%%%%%%%%%%%%%%%%%%%%%%%%%%%%%%%%%%%%%%%%%%%%%%%%%%%%%%%%%%%%%%%%%%%%%%%%%%%%%
%
%
%
The Casimir effect is a consequence of the distorted vacuum fluctuation spectrum of quantized fields in bounded domains or spaces with non-trivial topologies  \cite{Mostepanenko, Milton}.
In Casimir's original calculation, the system under consideration is the electromagnetic vacuum \cite{Casimir_1948}. 
Imposing Dirichlet boundary conditions along one spatial direction by confining the system between two (idealized) parallel plates causes a change in the (infinite) vacuum energy-density.
The variation of the vacuum energy-density with respect to the plate separation 
is called Casimir pressure, and, after renormalization, yields a finite 
expression for an attractive interaction energy per unit area between the plates.
The electromagnetic Casimir force, caused by quantum fluctuations of the 
electromagnetic vacuum, is varying as $\hbar c/L^4$ with the plate separation 
$L$ and the speed of light $c$.
It has been measured in a number of experiments using various experimental settings \cite[sec.~6]{Bordag_2001}.
The comparison between quantum vacuum experiments like these and the predictions 
of different theoretical models provides  possibility to test fundamental 
physics (like higher dimensions or additional interactions), in much the same way as accelerator experiments in high-energy physics do at the other end of the energy scale \cite{Gies_2008, Mostepanenko_2003}.

We consider here, instead of the electromagnetic vacuum, a weakly interacting 
Bose-Einstein-Condensate {(BEC)} at zero temperature, and expect, in a 
similar manner, the quantum fluctuations on top of the ground state of 
the {BEC} to give rise to observable Casimir forces:  
Within the Bogoliubov approximation, the excited states of a {BEC} can be treated as quasi-particles characterized by the dispersion relation 
\be
E(k) = \hbar c \sqrt{k^2 ( 1 + k^2 \zeta^2 )}
\label{eqn:bogo1}
\; ,
\ee
that behaves linear for small momenta, with the 
`sound velocity' $c = \hbar / (2 m \zeta)$ being inverse to the healing length 
$\zeta$. 
In \pr{eqn:bogo1}, the wave-number $1/\zeta$ characterizes the transition between the linear (phonon) and the quadratic (free-particle) regimes \cite{Stringari}.
$\zeta$ is also related to the $s$-wave scattering length $a$ of the atoms and to the {BEC} density $n$ via 
$\zeta = 1/(4 \sqrt{\pi n a}) $.
For small momenta, the quasi-particles (phonons) propagate in the same 
way as the massless electromagnetic field, except for the propagation velocity being different.
Hence, the zero temperature quantum fluctuations in a spatially confined {BEC} can be expected to result in an observable Casimir force. 

Different scenarios for 
Casimir forces in {BEC}s have been analyzed by previous work:
for the parallel plate geometry, an asymptotic expansion of the Casimir 
force has been calculated in \cite{Edery_2006}, the small expansion parameter being the ratio between healing length and plate separation.
In the leading order, it reproduces exactly the same $\hbar c/L^4$ 
behavior as in the electromagnetic vacuum. The next order 
corrections scale with the ratio $\zeta / L$.
Replacing the perfectly reflecting plates by impurities embedded in a 
quantum liquid, Casimir forces between these impurities 
have been calculated in \cite{Recati_2005, Klein_2005} as a function of the 
impurity-liquid coupling.
If the impurities are realized by atoms which, in a certain internal state, 
interact with the atoms of the quantum liquid through $s$-wave scattering,
the Casimir interaction should be detectable as a shift of spectral 
lines that depends on
the distance between the impurities \cite{Klein_2005}.
In the limit of an infinitely strong impurity-liquid coupling, the result for the 
(one-dimensional) parallel plate scenario was recovered \cite{Recati_2005}; 
for a weak coupling, however, 
the interaction between the impurities vanishes exponentially with the impurity separation on a scale set by the healing length.
For the ideal (i.e.\xspace non-interacting) Bose gas, it was found that there is no Casimir force at all between impurities of arbitrary interaction strength, including the idealized parallel plate scenario \cite[appendix~C]{Recati_2005}.
This is consistent with the quite general method of \cite{Bachmann_2008}, 
which is mapping (polynomial) dispersion relations to Casimir forces in 
the parallel plate geometry: this method shows that media with quadratic dispersion 
relations, and hence the ideal Bose gas, do not give rise to any zero-temperature 
Casimir forces. A non-vanishing Casimir force in 
the ideal Bose gas can arise due to \emph{thermal} fluctuations, as calculated 
in \cite{Biswas_2007,Martin_2006}.
All these forces are small but finite observable quantities, which---if experimentally confirmed---would provide direct evidence of the quantum fluctuations in weakly interacting {BEC}s.

The system under consideration in the present article is a homogeneous, weakly-interacting dilute {BEC} at zero temperature, confined to a parallel plate geometry with periodic boundary conditions in one of the three spatial dimensions.
The condition of diluteness can be formulated as $n |a|^3 \ll 1$, where $a$ is the $s$-wave scattering length and $n=N/V$ is the density, $N$ being the total particle number and $V$ the volume of the gas.
The quantity  $n |a|^3$ is usually called the gas parameter.
The perturbative calculation of typical properties of such a {BEC}, like the ground state energy or the depletion of the condensate, is then essentially an expansion in the gas parameter \cite{Stringari}. 

For this system, we will give a re-normalized expression for the Casimir energy-density per unit area. 
Our expression has the form of an integral over a `density of modes' 
$\rho( x )$ times the Bose distribution function:
\be
\ec = \int_0^{\infty} \frac{dx \,\rho(x)}{e^{2 \pi x} -1}
\; ,
\label{eq:Casimir-as-Bose-intgral}
\ee
where $\rho( x )$ has a simple analytic form (see \pr{eqn:rho_an}), and correctly describes the vanishing of the Casimir force in the limit of the interaction strength going to zero.
The possibility to express the zero temperature Casimir energy in the above form, resembling the density of states of a bosonic system at finite temperature, is connected to a topological analogy between our parallel plate scenario and finite temperature field theory:
In the parallel plate geometry, one spatial coordinate of the field is 
subject to periodic boundary conditions, while in finite temperature field 
theory, the imaginary time coordinate is subject to a similar periodicity 
condition. This analogy has been pointed out some time ago in 
 \cite{Toms_1980_a}; it does not carry over, however, to Casimir 
calculations for non-linear dispersion relations.
The periodic boundary conditions for the {BEC} have mainly been chosen
because they make the relation to the finite temperature case 
particularly evident. 
In experiments, periodic boundary conditions can be realized in toroidal 
traps, but they also appear in optical lattices.
If the boundaries are taken as real physical plates, the perfect mirror scenario 
(i.e.\xspace Dirichlet boundary conditions as discussed in
\cite{Bachmann_2008, Edery_2006_b, Biswas_2007})
is closer to an experimentally realizable situation. 
When Dirichlet boundary conditions are imposed on the fluctuations on
top of the {BEC} ground state, the expansion of $\ec$ for moderate
interaction shows in the leading term again the same behaviour as a
massless scalar field propagating at the speed of sound $c$, but with
a different numerical prefactor.

The article is organized as follows:
In \pr{sec:free}, we briefly recall some well-known perturbative expressions for the free energy, chemical potential and ground state energy of a dilute, weakly interacting {BEC} at zero temperature.
Working from the formula for the free energy, we derive in \pr{sec:casimir} the expression for the Casimir energy. 
In the limit $\zeta / L \ll 1$,  the function $\rho$ reproduces the  
power-series expression for the Casimir energy derived in \cite{Edery_2006}, 
as is shown in \pr{subsec:rho_expansion}.
In \pr{subsec:toms}, we relate the above mentioned analogy to the finite temperature case in some detail, as this topic does not seem to get overly much attention in the recent literature.
Finally, \pr{subsec:limit} deals with the behavior of the function $\rho$ 
in the limit of the effective coupling constant between the particles going to zero.
Here, our expression $\ec$  correctly describes the vanishing of the Casimir 
force. We comment on the failure 
of the large-distance expansion to describe the non-interacting Bose gas.
The case of Dirichlet boundary conditions is briefly discussed in the 
Appendix.

%
%%%%%%%%%%%%%%%%%%%%%%%%%%%%%%%%%%%%%%%%%%%%%%%%%%%%%%%%%%%%%%%%%%%%%%%%%%%%%%
\section{Free energy in a weakly interacting, dilute {BEC} at $T=0$}
\label{sec:free}
%%%%%%%%%%%%%%%%%%%%%%%%%%%%%%%%%%%%%%%%%%%%%%%%%%%%%%%%%%%%%%%%%%%%%%%%%%%%%%
%
In the one-loop approximation, the (unrenormalized) free energy density of a weakly interacting {BEC} at $T=0$ (in three spatial dimensions) is given by  
\be
\mc F 	% = - \frac{1}{V \beta} \, \ln [\mc Z ] 
	= \mc{F}_0 + \mc {F}_1  
\label{eqn:F1}
\ee
with the mean-field and one-loop contributions \cite{Andersen_2004}:
\bea
\mc{F}_0 	&=&  -\frac{\mu^2}{2 g}
\label{eqn:F1a} 
\\
\mc {F}_1	&=& \frac{1}{2} \, \dint k 3  \, \sqrt{k^2(k^2 + 2 \mu)} + \Delta_1 \mc F 
\, .
\label{eqn:F1b}
\eea
Here and in the following, we  work in units where $2m = \hbar = k_B 
= 1$.
In the above expression, 
$\mu$ is the chemical potential (which, in \pr{subsec:rho_expansion}, will be 
connected to the speed of sound in the medium), 
and $g$ is the effective coupling constant that can, for low energy collisions 
in a dilute medium, be identified with the $s$-wave scattering length $a$: 
\xspace$g = 8 \pi a$.  
The term $\Delta_1 \mc F$ represents the one-loop counter-term needed to render $\mc {F}_1$ finite.
If  the integral in \pr{eqn:F1b} is regularized with a momentum cut-off in the ultraviolet,  the linear, cubic and quintic UV-divergences in $\mc {F}_1$ can be absorbed by re-normalizing $g$, $\mu$ and the vacuum energy respectively.
If dimensional regularization is used in \pr{eqn:F1b}, no counter-term is needed at this level of perturbation theory.

After renormalization, the chemical potential $\mu$ can be obtained from \pr{eqn:F1}  by inverting
\be
n(\mu) = - \frac{\p \mc F}{\p \mu} \; ,
\ee
which yields
\be
\mu(n) = g n \bigl\{ 1 + \landau \bigl( \sqrt{n a^3}  \bigr) \bigr\} \; . 
\label{eqn:mu_def}
\ee
Reinserting $\mu$ into \pr{eqn:F1} will then reproduce the classical result for the leading quantum corrections to the ground state energy density $\mathcal E$ of a hard-sphere Bose gas, which was first derived by Lee, Huang and Yang in \cite{Lee_Yang_Huang_1957}:
\be
\mc{E} 	= \frac{g}{2} \, n^2 \, 
	\biggl\{
		1 \, + \, \frac{128}{15 \sqrt{\pi}} \, \bigl( n a^3 \bigr)^{1/2} \, + \, \landau \bigl( n a^3 \ln[n a^3]  \bigr)
	\biggr\}
\ee
%
%
%%%%%%%%%%%%%%%%%%%%%%%%%%%%%%%%%%%%%%%%%%%%%%%%%%%%%%%%%%%%%%%%%%%%%%%%%%%%%%
\section{The Casimir contribution as an integral over a density-function}
\label{sec:casimir}
%%%%%%%%%%%%%%%%%%%%%%%%%%%%%%%%%%%%%%%%%%%%%%%%%%%%%%%%%%%%%%%%%%%%%%%%%%%%%%
%
Now, in order to describe 
a {BEC} between a pair of parallel plates separated by a finite distance $L$ (with periodic boundary conditions), we have to quantize the momentum component perpendicular to the plates:
\be
k^2  	\; \to \;	k^2 + \omega_n^2, \quad \omega_n = \frac {2 \pi} L \,  n, \quad n \in \mathbb{Z}
\label{eqn:quant1}
\ee
Correspondingly, the momentum integration perpendicular to the plates is replaced by a discrete sum:
\be
\dint k 3 \; \to \; \sum_{n=-\infty}^{\infty} \, \dint k {2}	
\label{eqn:quant2}
\ee
The area of the plates is taken to be $L_1 L_2$ with $L_1,L_2 \gg L$, so the system now inhabits the volume $\overline{V} = L_1 L_2 L$.
The mean field contribution $\mc{F}_0$ (see \pr{eqn:F1a}) depends only trivially on the new boundary conditions, with the volume $\overline{V}$ entering through $\mu$ in \pr{eqn:mu_def}.

The Casimir energy $\ec$ of the BEC is related to the free energy 
at the 
one-loop level. We are interested in its change per unit area
that is due to the introduction of the boundary conditions:
\be
\overline{\mc{F}}_1 \; = \; L \mc{F}_1 \bigl|_{V = \overline{V}} \; + \; \ec
\label{eqn:F1d}
\ee
where the first term gives the (one-loop contribution to) the free
energy in a homogeneous system.
After applying the substitutions \pr{eqn:quant1} and \pr{eqn:quant2}
to \pr{eqn:F1b}, we are left with the following expression for $\overline{\mc{F}}_1$, now describing the leading quantum corrections to the free energy of a {BEC} confined between parallel plates:
\be
\overline{\mc{F}}_1 = \frac{1}{2 \, \lb^2} \, 
	    	\sum_{n=-\infty}^{\infty} \, \dint k 2
            	\sqrt{ 
		\bigl[\bigl( \lb \, k \bigr)^2 + n^2  \bigr] 
		\, \bigl[M(k)^2  + n^2  \bigr] }
\; ,
\label{eqn:F1c}
\ee
where we have used the abbreviations
\be
M(k,L,\zeta) 	= \lb \, \sqrt{k^2 + 1/\zb^2} 
\; ,
\label{eqn:M2_def}
\ee
with $\lb = L/ (2 \pi)$ and $\zb = 1/\sqrt{2 \mu}$.
The summation over $n$ can be converted into two integrals by using the 
Abel-Plana formula in the form (see \cite{Saharian_2007})
\be
\sum_{n=0}^\infty f(n) =
	\int_0^\infty dx\, f(x)
	+ \frac 1 2 \, f(0)
	+i \,\int_0^\infty dx \, \frac {f(ix) - f(-ix)}{e^{2 \pi x} -1}
\label{eqn:ap}
\ee
Application of \pr{eqn:ap} splits \pr{eqn:F1c} into the two terms
written in \pr{eqn:F1d}, as the $f(0)$--term cancels out.
The first term, $L \mc{F}_1 \bigl|_{V = \overline{V}}$,
is divergent, but of the same form as the one-loop free energy 
\pr{eqn:F1b} itself. Hence, it can be re-normalized as described above. 

The term $\ec$ in \pr{eqn:F1d}, an energy per area, describes the
effects induced by restricting the periodicity in one spatial
dimension to be much smaller than the remaining ones, i.e.,\xspace
confining the system to a volume $\overline{V}$ with one `short' side
$L$.
$\ec$ is convergent for any finite $L$, strictly
negative, and goes to zero for $L \to \infty$:
\be
\fl
\ec	= - \frac{2}{\lb^2} \, \dint k 2 \int_{\lb k}^{M(k)}  
     dx\, \frac{\bigl[x^2 - \bigl(\lb \, k \bigr)^2 \bigr]^{1/2} \,
     \bigl[M(k)^2 -x^2 \bigr]^{1/2} }{e^{2 \pi x} -1}
\label{eqn:cas1}
\ee
The integration domain is sketched in \pr{fig:wedge-domain} (the shaded regions).
%
%
%
%
%%%%%%%%%%%%%%%%%%%%%%%%%%%%%%%%%%%%%%%%%%%%%%%%%%%%%%%%%%%%%%%%%%%%%%%%%%%%%%
\begin{figure}
\centering
\subfigure[Area of integration in \pr{eqn:cas1}.]
{
\includegraphics[width=0.39\textwidth]{figure1.epsi}\label{fig:wedge-domain}
}
\hspace{0.02\textwidth}
\subfigure[`Density of states' function $\rho( x )$.]
{
\includegraphics[width=0.39\textwidth]{figure2.epsi}\label{fig:rho_plot}
}
\caption{\\
\noindent
Fig.~\ref{fig:wedge-domain}: 
The upper curve shows the function $M(k,L,\zeta)$ from \pr{eqn:M2_def}, the lower one is 
$\lb k \equiv L k / (2\pi)$.
The two shaded regions make up the total area of integration in \pr{eqn:cas1}.
\\
\noindent
Fig.~\ref{fig:rho_plot}: The exact `density of states' function $\rho( x )$ from
\pr{eqn:rho_an} (solid line) and its small-argument expansion
\pr{eqn:rho_ex1}) (broken line). 
}
\end{figure}
%%%%%%%%%%%%%%%%%%%%%%%%%%%%%%%%%%%%%%%%%%%%%%%%%%%%%%%%%%%%%%%%%%%%%%%%%%%%%%
%
%
This domain as well as the form of the integrand in \pr{eqn:cas1} are due to the branch-points of the integrand in \pr{eqn:F1c}:
In order to stay clear of the branch-cuts, we have evaluated the last term  in the Abel-Plana formula \pr{eqn:ap} along an integration contour slightly to the right of the imaginary axis.
Inserting the function $f$ given by \pr{eqn:F1c},
\be
f(x) = \sqrt{(x^2 + \lb^2 \, k^2)(x^2 + M^2)} \; ,
\ee
the numerator in the last term of \pr{eqn:ap} evaluates to
\be
\fl
i \, \bigl( f(i x + 0) - f(-i x + 0)  \bigr) = -2 \, \sqrt{(x^2 - \lb^2 \, k^2)(M^2 - x^2)} 
\,, \qquad
\lb \, k < x  < M 
\; ,
\ee
being zero everywhere else along the contour of integration (see \cite[sec.~2.2]{Mostepanenko}).
By changing the order of integration in \pr{eqn:cas1}, the 
$k$-integral can be performed, and we find the 
expression \pr{eq:Casimir-as-Bose-intgral} for the Casimir energy:
\be
\ec	=  \int_0^{\infty} \frac{dx \, \rho (x, \lb)}{e^{2 \pi x} -1}
\label{eqn:DOS1}
\ee
In the form of \pr{eqn:DOS1}, $\ec$ is expressed as an integral over a
`density of states' {(DOS)} for a bosonic system, as setting the
integration variable $x = \beta \omega /(2\pi)$ will reproduce the
Boltzmann factor in the denominator.

The `mode density' $\rho( x )$ is obtained by 
integrating separately over the lower triangular region in 
\pr{fig:wedge-domain} (dark gray) and the `hyperbolic tail' (light gray).
In terms of the dimensionless variable
$\eta = (\zb /\lb) \, x$, we have
\be
\fl
\rho(x,\lb) =\cases{
-\frac{1}{8 \pi \zb^4} \,	
	\bigl\{
	\arcsin ( \eta)   
	 - \eta \, \bigl( 1 - 2 \, \eta^2  \bigr) \, \sqrt{1  - \eta^2}
	\bigr\}  
&for $0 \le x <  \lb / \zb$	
\\
- \frac{1}{16\,\zb^4}
&for  $x \ge \lb / \zb$	
}
\label{eqn:rho_an}
\ee
Above $x = \lb / \zb$, the function $\rho(x)$ changes into a constant independent 
of $x$ (see the solid line in \pr{fig:rho_plot}). The Casimir energy $\ec$ as a 
function of the normalized distance $\lb/\zb$ is shown in 
\pr{fig:E_plot}, obtained by numerically integrating \pr{eqn:DOS1}.

To provide a cross-check for the above results, we will in the following section show that the function $\rho(x)$  reproduces the asymptotic expansion for the Casimir energy  given in \cite{Edery_2006, Roberts_2005}, which is valid for small values of  the parameter $\zeta / L$.
In \pr{subsec:limit} we then show that the  formula for $\rho(x)$ in \pr{eqn:rho_an} also yields correct results in the opposite limit of the ideal {BEC}, i.e.\xspace $\zeta \to \infty$. 
%
%
%
%%%%%%%%%%%%%%%%%%%%%%%%%%%%%%%%%%%%%%%%%%%%%%%%%%%%%%%%%%%%%%%%%%%%%%%%%%%%%%
\begin{figure}
\centering
\includegraphics[width=0.8\textwidth]{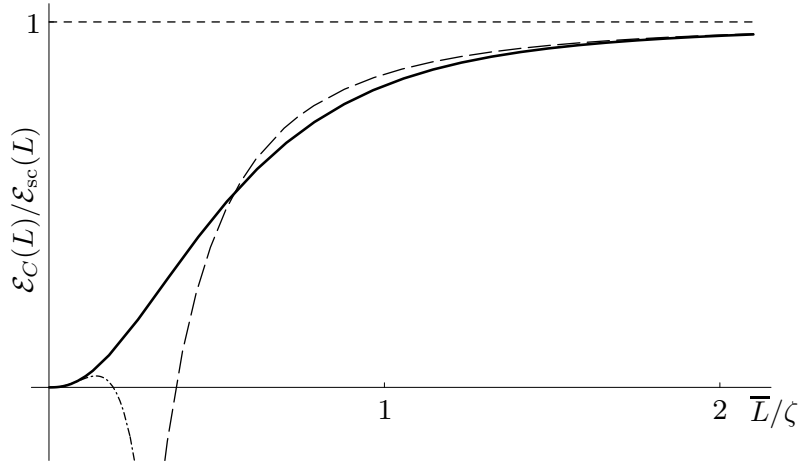}
\caption{Casimir energy per unit area as a function of the
plate separation $\lb = L/(2\pi)$, normalized to the healing length $\zeta$. Recall 
that $\zeta = \frac {\hbar} 2 \sqrt{1/(g n m)}$ with $g$ the effective interaction 
constant and $n$ the BEC density.   The Casimir energy has been
normalized to its value $\esc ( L ) = - (\pi^2/90)
\hbar c / L^3$ for a massless
scalar field propagating with a velocity $c = \hbar/ (2 m \zb)$; this limit is 
approached at large distance (\dashed).  The broken curve
(\longbroken) shows the asymptotic expansion for $L/\zeta \gg 1$
\pr{eqn:cas_ex}, while the dotted curve (\chain) shows 
the opposite limit $L/\zeta \to 0$ (non-interacting Bose gas), 
of eqn.~\pr{eqn:E_nonint}.  The full black curve is obtained by numerically
evaluating the integral in \pr{eqn:DOS1} with the mode density $\rho(
x )$ of \pr{eqn:rho_an}.  It smoothly describes the dependence of
$\ec$ on the interaction strength in the regime where both of the
asymptotic expansions diverge.}
\label{fig:E_plot}
\end{figure}
%%%%%%%%%%%%%%%%%%%%%%%%%%%%%%%%%%%%%%%%%%%%%%%%%%%%%%%%%%%%%%%%%%%%%%%%%%%%%%
%
%
%
%
%
%%%%%%%%%%%%%%%%%%%%%%%%%%%%%%%%%%%%%%%%%%%%%%%%%%%%%%%%%%%%%%%%%%%%%%%%%%%%%%
\subsection{Expansion at large distance and moderate interactions}
\label{subsec:rho_expansion}
%%%%%%%%%%%%%%%%%%%%%%%%%%%%%%%%%%%%%%%%%%%%%%%%%%%%%%%%%%%%%%%%%%%%%%%%%%%%%%
%
%
In this section, we will assume the plate separation to be much
greater than the healing length of the {BEC}, i.e.,\xspace the ratio
$\zeta /L \ll 1$ can be treated as a small parameter. Note that this 
limit cannot describe a strongly interacting Bose gas in the proper 
sense, since this would lead to the break-down of the one-loop 
approximation at the basis of our approach.

As can be seen in \pr{fig:rho_plot}, the kink in the function $\rho(x)$  
happens at the large value $x = \lb/\zb \gg 1$ in our limit.
In evaluating the Casimir energy with \pr{eqn:DOS1}, large values of $x$, and hence the behavior of $\rho(x)$ after the kink, get exponentially suppressed by the denominator.  
Thus we can approximate the function $\rho(x)$ by expanding its small-$x$ part 
as a power-series around $x=0$:
\be
\rho (x) 	= - \frac{1}{\lb^4} 	
		\biggl\{
		\frac{1}{\pi} \frac{x^3}{3} \, \biggl( \frac{\zb}{\lb} \biggr)^{-1} 
		 - \, \frac{1}{\pi} \frac{x^5}{10} \, \frac{\zb}{\lb} \,
		+ \, \landau\bigl( (\zeta/L)^3 \bigr) 
		\biggr\}
\label{eqn:rho_ex1}
\ee
The integrals in 
the expression \pr{eqn:DOS1} for the Casimir energy can be 
performed explicitly.\footnote{
We use the identities
\begin{equation*}
\int_0^\infty dt \,  \frac {t^{2 r -1}} {e^{2 \pi t} - 1} 
        = \frac{ \Gamma(2 r) \, \zeta( {2 r} ) }{(2 \pi)^{2 r}}
\label{eqn:integral_formula}
\end{equation*}
in terms of the Gamma- and Zeta-functions.}
Upon setting $c:= 1 / \zb$ (the speed of sound in the medium), we recover the result derived in \cite{Edery_2006}: 
\be
\ec 	= -\frac{\pi^2}{90} \, \frac{c}{L^3} 
			\;+\; \frac{2 \pi^4}{315} \, \frac{c \, \zeta^2}{L^5} 
			\;+\; \landau\bigl( \zeta^3/L^7 \bigr)
\label{eqn:cas_ex}
\ee
The leading term ${\mc E}_{\rm sc} = -\pi^2 c/(90 L^3)$ is, as noted
in \cite{Edery_2006,Edery_2006_b}, equal to the Casimir energy (per 
unit area)
of a massless scalar field with propagation velocity $c$ confined
between two parallel plates with periodic boundary conditions
\cite{Svaiter_1991}.  Its presence is a manifestation of the
Goldstone-theorem, the long wavelength part of the Bogoliubov spectrum
representing the gap-less Goldstone modes \cite{Hugenholtz_1959}.  The
next-to-leading term in \pr{eqn:cas_ex} is in \cite{Edery_2006}
referred to as the \emph{Bogoliubov correction} due to the
non-linearity of the dispersion.
Indeed, if we express the Bogoliubov dispersion relation \pr{eqn:bogo1} as a power series around $k=0$ and put the first few terms into the expression for the Casimir force derived in \cite[eqn.~(16)]{Bachmann_2008}, the leading term in the same manner reproduces the result for the scalar field (in one dimension), followed by terms that are smaller in magnitude and of opposite sign.
As can be seen in \pr{fig:E_plot}, the contribution of these
corrections is getting smaller as the ratio between plate separation
and healing length increases, leaving only the dominant scalar term
$\esc$ depicted by the horizontal dashed line in
\pr{fig:E_plot}.

When finally calculating the Casimir pressure from \pr{eqn:cas_ex}, one has to consider that, as the number of particles is held constant, the derivative of the speed of sound with respect to $L$ is not zero,
as discussed in \cite{Edery_2006}.
%
%%%%%%%%%%%%%%%%%%%%%%%%%%%%%%%%%%%%%%%%%%%%%%%%%%%%%%%%%%%%%%%%%%%%%%%%%%%%%%
\subsection{A formal analogy to finite temperature systems}
\label{subsec:toms}
%%%%%%%%%%%%%%%%%%%%%%%%%%%%%%%%%%%%%%%%%%%%%%%%%%%%%%%%%%%%%%%%%%%%%%%%%%%%%%
%
%
As already mentioned, our expression \pr{eqn:DOS1} for $\ec$ at $T=0$ formally resembles the {DOS} for a bosonic system at finite temperature.
This can be understood by recalling that a finite temperature
system can be described in imaginary time, combined with periodic boundary
conditions with period $\beta$.  The same topology is realized in the
parallel plate system at $T=0$, when one spatial dimension is subject to
periodic boundary conditions with period $L$ (\cite{Toms_1980_a}, see
also \cite {Hawking_1977}).

Let us briefly re-phrase the argument of \cite{Toms_1980_a} 
within our notation.
The canonical partition function $\mc{Z}$ for a system at
temperature $T = 1 / \beta$ in $D$ spatial dimensions is usually
expressed as the path integral
\be
\mc{Z} \approx \oint d[\Phi] \, \exp \biggl[\int_0^\beta d\tau \, \int d^D x \, \mc{L}(\Phi)\biggr]
\; ,
\label{eqn:toms_Z}
\ee
where $\mc L$ is the Lagrangian (a scalar functional of the field $\Phi$), and the field is constrained in such a way that  $\Phi({\bf x},0) =  \Phi({\bf x},\beta)$.
If we take  $\mc{L} = (c^2/2) \, \p_\mu \Phi \, \p^\mu \Phi$, the path-integral in \pr{eqn:toms_Z} can, after Fourier expansion of the field, be evaluated to yield \cite{Bernard_1974, Kapusta}
\be
\fl
\ln [\mc {Z}] 	= - V \, \frac{1}{2} 
		\sum_n \int \frac{d^D k}{(2 \pi)^D} \, \ln \bigl[ \omega_n^2 + 
		\omega( k )^2  \bigr] \,
		, \quad \omega_n = \frac {2 \pi}{\beta} n \,
		, \quad \omega( k ) = c \, |\vec{k}|
\label{eqn:toms_logz_a}
\; .
\ee
Here, $\omega_n$ with $n = 0, \pm 1, \pm 2, \dots$ are the Matsubara-frequencies 
due to the periodicity condition in \pr{eqn:toms_Z}, and $\omega_k$  with the continuous parameter $k$ is the dispersion relation of the massless scalar field.
Note the similarity to \pr{eqn:quant1}, where we had periodic boundary conditions not in imaginary time but in one spatial dimension.

The sum over $n$ in \pr{eqn:toms_logz_a} is usually evaluated by multiplying 
with a factor $\frac 1 2 \beta \, \cot (\frac 1 2 \beta \omega)$, which has 
poles of residue 1 at $\omega=2 \pi n / \beta$, and then integrating over a contour in the complex $\omega$-plane which includes all the poles \cite[sec.~3.4]{Kapusta}.
This technique is actually the same that is used in \cite{Saharian_2007} 
to prove the Abel-Plana formula \pr{eqn:ap}. 
The well known result is 
\be
\ln [\mc Z]  =  \ln [\mc Z]\bigg|_{\beta \to \infty}
		- \, V \, \int \frac{d^D k}{(2 \pi)^D} \, \ln \bigl[ 1 - e^{- \beta \, \omega(k)}  \bigr]
\;,
\label{eqn:toms_logz_bose}
\ee
where we have already subtracted the zero-point fluctuations.
After integrating by parts and employing the 
integral-identity from the footnote on page \pageref{eqn:integral_formula}, 
we obtain ($\Omega_{D}$ is the volume of the unit sphere in $D$ 
dimensions)
\be
\ln [\mc Z] 	
	= 
	- V \beta c \, \frac{\Omega_D / D}{(2 \pi)^D} 
	\int_0^\infty \, \frac{dk \, k^D} {e^{\beta \, c k} - 1}
\; .
\ee
Note that, upon setting $k = (2 \pi/\beta c)\,x$, the above expression has the same form as $\ec$ in \pr{eqn:DOS1}, with $\rho(x)$ being proportional to $x^D$.
Now, let the system  inhabit a volume $\overline{V} = L_1 L_2 L$.
With $D=3$ we get for the free energy per unit area 
\be
\frac{F}{L_1 L_2} = - \frac{\pi^2}{90} \, \frac{L}{\beta \, (\beta c)^3}
\label{eqn:toms_logZ_3d}
\; .
\ee
Tentatively exchanging $\beta c$ with $L$ in \pr{eqn:toms_logZ_3d} will reproduce the Casimir energy for a massless scalar field, i.e.\xspace the first term in \pr{eqn:cas_ex} \cite{Toms_1980_a}: 
\be
\frac{F}{L_1 L_2}
	\; \overset{\beta c \leftrightarrow L}{\longrightarrow} \; 
	-\frac{\pi^2}{90} \, \frac{c}{L^3} = 	\esc
\label{eqn:toms_trick}
\ee
So, we have seen that the zero temperature Casimir energy of a
massless scalar field confined between two parallel plates can be
obtained by a simple exchange of variables, once we know the thermal
contribution to the free energy density of that field.
Unfortunately, this simple mapping does not carry over to fields 
characterized by nonlinear dispersion relations: 
The temperature dependent part of the free energy for a {BEC} is (in
the Bogoliubov approximation) still given by \pr{eqn:toms_logz_bose}
with the dispersion now being $\omega(k) = c k \, (k^2 + 1/\zb^2)^{1/2}$
\cite{Andersen_2004}. 
But a simple interchange of $\beta c$ with $L$ in
\pr{eqn:toms_logz_bose} will no longer yield $\ec$, because the
discretized Matsubara frequencies $\omega_n$ 
(that become a discretized momentum)
always enter in \pr{eqn:toms_logz_a} in the same way 
as a spatial momentum component enters
in a \emph{linear} dispersion relation. It thus seems very difficult 
to mimic the fully nonlinear behavior of the dispersion relation.
%
%
%%%%%%%%%%%%%%%%%%%%%%%%%%%%%%%%%%%%%%%%%%%%%%%%%%%%%%%%%%%%%%%%%%%%%%%%%%%%%%
\subsection{The non-interacting limit}
\label{subsec:limit}
%%%%%%%%%%%%%%%%%%%%%%%%%%%%%%%%%%%%%%%%%%%%%%%%%%%%%%%%%%%%%%%%%%%%%%%%%%%%%%
%
%
The transition from the weakly interacting {BEC} (here described at the 
one-loop level only) to the ideal {BEC} should be accomplished by letting 
the effective coupling $g  = 8 \pi a$ (see \pr{sec:free}) go to zero, 
corresponding to $\zeta \to \infty$.
The Casimir energy is expected to vanish in this limit, as shown in 
 \cite{Recati_2005,Bachmann_2008}.

The series expansion \pr{eqn:rho_ex1} was constructed for $\zeta / L \ll 1$, which is a physically reasonable assumption for finite L and a weak but finite interaction.
But with the effective coupling strength $g \to 0$, the chemical potential $\mu$  in \pr{eqn:mu_def} will vanish, too, and the healing length $\zeta$ will diverge. 
As $L$ is kept finite, the non-interacting {BEC} is hence described by
the limit $L/\zeta \to 0$, which is the opposite to the case
considered in \pr{subsec:rho_expansion}. The asymptotic form of 
\pr{eqn:rho_ex1} for the mode density 
$\rho(x)$---as well as the Casimir energy in
\pr{eqn:cas_ex}---diverges in the limit of zero interaction
strength. Our calculation of the exact mode density $\rho(x)$ 
suggests that this divergence is due to a branch point in the complex 
$x$-plane that moves towards $x = 0$ and makes the power series 
expansion behind \pr{eqn:cas_ex} break down. We show here that the exact 
mode density~\pr{eqn:rho_an} leads to a Casimir energy that smoothly
vanishes with the interaction strength (see \pr{fig:E_plot}).

To examine the asymptotic behavior of $\ec$ for $\lb / \zb \ll 1$, we again start from \pr{eqn:DOS1} and \pr{eqn:rho_an}, separately treating the behavior of $\rho(x)$ to the left and the right of the kink at $\lb/\zb$:
\bea
\ec 	
	&=&
	\int_0^{\lb / \zb} \frac{dx \, \rho(x)}{e^{2 \pi x} - 1}
	- \frac{1}{16 \zb^4} \, \int_{\lb/\zb}^{\infty}\frac{dx}{e^{2 \pi x} - 1}
\\	
	&=&
	\mc{E}_1 + \mc{E}_2
\label{eqn:E_nonint1}	
\eea
As the upper limit of the integration in $\mc{E}_1$ is going to zero, we can replace $\rho(x)$ by the first term in the expansion of \pr{eqn:rho_ex1}, and Taylor-expand the denominator about $x=0$.
After integrating over this expansion, $\mc{E}_1$ will yield terms of $\landau\bigl( (\lb / \zb)^4 \bigr)$.
Integrating $\mc{E}_2$, from the lower border of the integral we get a contribution
\be
\mc{E}_2  	= \frac{\pi^3}{2 L^4} \, \biggl( \frac{\lb}{\zb}\biggr)^4 \, \ln \bigl[ e^{2\pi \, (\lb / \zb)} -1  \bigr]
\; . 
\label{eqn:E2_nonint}
\ee
Altogether, we find for the behavior of $\ec$ in the non-interacting limit
\be
\fl
\ec = -\frac{1}{L^4} \, \biggl( \frac{\lb}{\zb}\biggr)^4 \, 
                        \biggl\{  
			\frac{4 \pi^2}{3}
			- \frac{\pi^3}{2}  \,  \ln \bigl[ e^{2\pi \, (\lb / \zb)} -1  \bigr]
      	 	        \biggr\}
			+ \landau\bigl( (\lb / \zb)^5 \bigr)
\; . 
\label{eqn:E_nonint}
\ee
Note that the leading order for $\zb \to \infty$ at fixed $L$ goes 
like $\zb^{-4} \ln \zb$. Conversely, at fixed $\zb$, a logarithmic 
divergence remains for $L \to 0$. 
The logarithmic term changes sign for $L / \zb > \ln[2] \approx 0.7$ and, for large values of $\lb / \zb$,  the above expression diverges, just as the expansion \pr{eqn:cas_ex} does for small values of $\lb / \zb$ (see \pr{fig:E_plot}).
Hence, \pr{eqn:E_nonint} and  \pr{eqn:cas_ex} provide two asymptotic expansions to $\ec$ for opposite limits, while the exact formula is given by \pr{eqn:DOS1} integrated over \pr{eqn:rho_an}. 
 
%
%
%%%%%%%%%%%%%%%%%%%%%%%%%%%%%%%%%%%%%%%%%%%%%%%%%%%%%%%%%%%%%%%%%%%%%%%%%%%%%%
\section{Summary}
\label{sec:summary }
%%%%%%%%%%%%%%%%%%%%%%%%%%%%%%%%%%%%%%%%%%%%%%%%%%%%%%%%%%%%%%%%%%%%%%%%%%%%%%
%
%
Starting from the free energy in a weakly-interacting dilute {BEC}, 
we derived a re-normalized expression for the `phononic' Casimir energy of the {BEC} confined at zero temperature to a parallel plate geometry with periodic boundary conditions.
Our formula for the Casimir energy (per unit plate area), \pr{eqn:DOS1}, has the form of an integral over a density function $\rho$ times the Bose distribution.
The function $\rho$ is given by a rather simple analytic expression in \pr{eqn:rho_an}.
In \pr{subsec:rho_expansion}, we provided a cross-check for our result by showing that a series expansion of $\rho$ in the parameter $L/\zeta \gg 1$ reproduces the asymptotic series for the Casimir energy derived in \cite{Edery_2006}.
There, the Euler-MacLaurin formula was used to extract the long wavelength behavior out of the UV-divergent sum over all Bogoliubov modes satisfying the boundary conditions.
This approach fails to reproduce the non-interacting limit.

As pointed out in \cite{Edery_2006_b,Edery_2006}, the Casimir energy of the weakly interacting {BEC} is,
due to the linear dispersion of its low lying excitations,
in the leading order determined by a term analogous to the Casimir energy of a massless scalar field propagating with vacuum velocity $ c = 1 / \zb$.
Our result displays this behavior in the regime of the weakly interacting {BEC} where the plate separation is much larger than the healing length, as can be seen in \pr{fig:E_plot}.
In addition, for $\zeta \to \infty$ with $L$ kept finite (the non-interacting limit), our result \pr{eqn:rho_an}  correctly describes the Casimir energy going to zero and displays the Casimir energy as a smoothly varying function of the interaction strength in the intermediate range.    
The subtleties of the asymptotic expansions illustrate the rich physical 
content behind the nonlinear dispersion relation of the Bogoliubov vacuum. 
%
%%%%%%%%%%%%%%%%%%%%%%%%%%%%%%%%%%%%%%%%%%%%%%%%%%%%%%%%%%%%%%%%%%%%%%%%%%%%
% ACHNOWLEDGMENTS
%%%%%%%%%%%%%%%%%%%%%%%%%%%%%%%%%%%%%%%%%%%%%%%%%%%%%%%%%%%%%%%%%%%%%%%%%%%%
\ack
This research was supported by \emph{Deutsche Forschungsgemeinschaft} (DFG), grant He~2849/3.
%
%
%
%
%
%
%
%%%%%%%%%%%%%%%%%%%%%%%%%%%%%%%%%%%%%%%%%%%%%%%%%%%%%%%%%%%%%%%%%%%%%%%%%%%%
% APPENDIX
%%%%%%%%%%%%%%%%%%%%%%%%%%%%%%%%%%%%%%%%%%%%%%%%%%%%%%%%%%%%%%%%%%%%%%%%%%%%
\appendix
\section*{Appendix: Dirichlet boundary-conditions}
\setcounter{section}{1}
Here we discuss how our results for the Casimir energy
get modified with Dirichlet boundary conditions instead of periodic ones:
taking the wavenumbers $\omega_n$ in \pr{eqn:quant1} as $\omega_n = n \pi / L$ with $n$ running from $1$ to infinity,
application of the Abel-Plana formula generates an additional divergent surface-term  $\mathcal{E}_{surf}$ in \pr{eqn:F1d}:
\be
\mathcal{E}_{surf} = 
	-\frac{1}{4} \, \int \frac{d^2 k}{(2 \pi)^2} \, \sqrt{k^2 (k^2 + 1/\zeta^2)}
\label{eqn:surf}
\ee
(The same term with opposite sign will appear for von Neumann boundary
conditions, similar to \cite{Martin_2006} where the ideal gas at
finite temperature is considered.)
As $\mathcal{E}_{surf}$ does not depend on $L$, it does not affect the Casimir energy.
Our main formula for $\ec$, \pr{eqn:DOS1}, 
is modified by an overall factor $\case 1 2$,
while in the mode density $\rho (x, \lb)$ (see \pr{eqn:rho_an})
the distance argument picks up an additional factor of two:
\be
\ec	=  \frac{1}{2} \,\int_0^{\infty} \frac{dx \, \rho (x, 2\lb)}{e^{2 \pi x} -1}
\label{eqn:DOS1_dir}
\ee
The asymptotic expansion for large plate separation and moderate interactions, 
as derived in \pr{subsec:rho_expansion},  now yields (compare with \pr{eqn:cas_ex}): 
\be
\ec 	= -\frac{\pi^2}{1440} \, \frac{1}{\zeta \, L^3} 
			\;+\; \frac{\pi^4}{10080} \, \frac{\zeta}{L^5} 
		        \;+\; \landau\bigl( \zeta^3/L^7 \bigr) 
\; , \qquad L / \zeta \gg 1.
\label{eqn:cas_ex_dir}
\ee
Again, the leading term is equal to the known result for the 
Casimir energy of a massless scalar field with propagation velocity $c=1/\zeta$ 
confined between two parallel ideal mirrors (see \cite[eqn.~(38)]{Edery_2006_b}).
For the non-interacting limit, as treated in \pr{subsec:limit}, we get the asymptotic expression
(compare with \pr{eqn:E_nonint})
\be
\fl
\ec = -\frac{1}{L^4} \, \biggl( \frac{\lb}{\zb}\biggr)^4 \, 
                        \biggl\{  
			\frac{4 \pi^2}{9}
			- \frac{\pi^3}{4}  \,  \ln \bigl[ e^{4\pi \, (\lb / \zb)} -1  \bigr]
      	 	        \biggr\}
			+ \landau\bigl( (\lb / \zb)^5 \bigr)
\; , \qquad L / \zeta \ll 1.
\label{eqn:E_nonint_dir}
\ee
This goes to zero as for periodic boundary conditions, with slightly 
different numerical factors. 

%\noindent
The divergent surface energy \pr{eqn:surf}
does not fit into the renormalization scheme for the bulk part of the free energy $\mathcal {F}_1$
as described after \pr{eqn:F1b}.
For a full treatment of the free energy of a {BEC} between perfect mirrors,
it is not sufficient to change only the wavenumbers $\omega_n$ in the 
Bogoliubov excitations,
because both the ground-state wave function and the excitations 
have to be zero on the boundaries.
A detailed analysis will be reported elsewhere.
%  
%
%
%
%
%
%
%
%%%%%%%%%%%%%%%%%%%%%%%%%%%%%%%%%%%%%%%%%%%%%%%%%%%%%%%%%%%%%%%%%%%%%%%%%%%%
% REFERENCES
%%%%%%%%%%%%%%%%%%%%%%%%%%%%%%%%%%%%%%%%%%%%%%%%%%%%%%%%%%%%%%%%%%%%%%%%%%%%  
%
\section*{References}
%\bibliographystyle{unsrt}	% 
%\bibliography{casimir}	        % (uses file casimir.bib)

%
%%%%%%%%%%%%%%%%%%%%%%%%%%%%%%%%%%%%%%%%%%%%%%%%%%%%%%%%%%%%%%%%%%%%%%%%%%%%
%%%%%%%%%%%%%%%%%%%%%%%%%%%%%%%%%%%%%%%%%%%%%%%%%%%%%%%%%%%%%%%%%%%%%%%%%%%%
%
\end{document}